\documentclass[aps,prl,twocolumn,showpacs,amsmath,amssymb,superscriptaddress]{revtex4-1}
\usepackage{graphicx}
\usepackage{dcolumn}
\usepackage{bm}
\usepackage{natbib}

\begin{document}

\title{Spectral origin of the colossal magnetodielectric effect in multiferroic DyMn$_2$O$_5$} 

\author{A. B. Sushkov}
\affiliation{Materials Research Science and Engineering Center and Center for Nanophysics and Advanced Materials, University of Maryland, College Park, Maryland 20742}
\thanks{sushkov@umd.edu}
\author{Ch.~Kant}
\author{M.~Schiebl }
\author{A.~M.~Shuvaev}
\author{Anna~Pimenov}
\author{Andrei~Pimenov}
\affiliation{Institute of Solid State Physics, Vienna University of Technology, 1040 Vienna, Austria}
\author{Bernd Lorenz}
\affiliation{Texas Center for Superconductivity and Department of Physics, University of Houston, Houston, Texas 77204-5002}
\author{S. Park}
\author{S-W. Cheong}
\affiliation{Rutgers Center for Emergent Materials and Department of Physics and Astronomy, Rutgers University, Piscataway, New Jersey 08854}
\author{Maxim Mostovoy}
\affiliation{Zernike Institute for Advanced Materials, University of Groningen, 9747 AG Groningen, The Netherlands}
\author{H. D. Drew}
\affiliation{Materials Research Science and Engineering Center and Center for Nanophysics and Advanced Materials, University of Maryland, College Park, Maryland 20742}

\date{\today}
\begin{abstract}
The origin of the colossal magnetodielectric (CMD) effect in DyMn$_2$O$_5$~\cite{Hur-Dy} has been an outstanding question in multiferroics. 
Here, we report the activation of the electric dipole mode at 4--5~cm$^{-1}$ in an applied magnetic field which fully accounts for the CMD effect. We examine two alternative explanations of this mode: a soft electromagnon and transitions between f-electron levels of Dy$^{3+}$ ions. The experimental and theoretical evidence supports the electromagnon origin of the CMD effect.
\end{abstract}

\pacs{78.20.Ls, 78.30.-j, 75.85.+t,76.30.Kg}
 
\maketitle
Large enhancement of the dielectric constant, $\varepsilon$, in an applied  magnetic field, called colossal magnetodielectric or giant magnetocapacitance effect, is one of the most spectacular phenomena occuring in  multiferroic materials with the magnetically-induced electric polarization. This effect was first observed in DyMnO$_3$, in which the electric polarization is induced by a cycloidal spiral ordering~\cite{Goto2004}. In an applied magnetic field the spiral plane and the electric polarization vector rotate through 90 degrees, which results in the $\sim$ 500~\% increase of the dielectric constant. The origin of this large dielectric susceptibility was traced back to the high mobility of the 90$^{\circ}$ magnetic domain walls separating the two multiferroic states with orthogonal electric polarizations, which makes possible to move these walls with an electric field~\cite{Kagawa2009}. A characteristic signature of this mechanism is the disappearance of the CMD effect at frequencies of the order of 10~MHz.

A comparable increase of dielectric constant in a magnetic field of 7$\div$8~T was found in the multiferroic DyMn$_2$O$_5$ \cite{Hur-Dy} and initially was also attributed to the relaxational dynamics of ferroelectric (FE) domain walls~\cite{GS-JETPL}. In this Letter, we report an investigatation of the spectral origin of the CMD effect in DyMn$_2$O$_5$ using the magnetic field and temperature dependences of its far infrared transmission spectra and measurements of the dielectric constant in MHz range. We show that in this material the CMD effect originates from  an emergent electric dipole (ED) mode in the THz frequency range (called here the X-mode) rather than kHz-MHz domain wall motion, as in the case of DyMnO$_3$.

We consider two possible candidates for this mode. One is the tantalizing possibility of an electromagnon - which is short for "magnon with electric dipole activity". 
Theoretically~\cite{Katsura-PRL-2007,us-JPCM-2008,Fang-EPL-2008,deSousa2008,Cao-JPCM-2012}, the electromagnon is a {\it linearly} coupled lattice (phonon) and magnetic (magnon) mode resulting in mode mixing:"repulsion" of the original  phonon and magnon frequencies as well as a partial exchange of their oscillator strengths.  
The electromagnon is distinct from the {\it nonlinear} dynamic coupling effects such as spin-phonon coupling, two-magnon coupling, ED-active two-magnon coupling, or Raman-active magnons.
Recently discovered electromagnons~\cite{Pimenov-Nature,us-PRL-2007,Kida-2009} are result of nonresonant magnon-phonon coupling at $q$=0 in contrast with the well known "anti-crossing" of magnon and phonon branches~\cite{Kittel-1958} that occur at $q$$\neq$0 and are detectable only by inelastic neutron scattering.
According to recent analysis of inelastic neutron and infrared data on YMn$_2$O$_5$ \cite{Kim-PRL-2011}, this no-rare-earth~\cite{Kenzelmann-view} multiferroic compound has three electromagnon modes: optical phason near 1~meV and a doublet at 2.5~meV.

An alternative possible origin of dielectric constant anomalies in rare earth oxides is associated with electric-dipole transitions between the $f$ levels of rare earth ions.  
In both $R$Mn$_2$O$_5$ and $R$MnO$_3$, the rare earth ion occupies a site of very low symmetry $C_s$. As a result, the crystal and ligand fields mix the $f$ levels with the levels of opposite parity, thus allowing the originally ED forbidden $f$-$f$ transitions to borrow some ED spectral weight from the ED-allowed electronic excitations. 
We note that the largest step-like anomaly of the dielectric constant is observed in the 1-2-5 manganites with Dy$^{3+}$ and Ho$^{3+}$~\cite{Hur-Dy}, which are known to have the lowest frequencies of the $f$-$f$ transitions among all rare earth ions.

The results of this work have been obtained on flux-grown single crystals~\cite{Hur-Dy}. 
Transmission measurements were performed on the Bomem DA8 Fourier-transform spectrometer in the frequency range 10$\div$240~cm$^{-1}$ and on the Mach-Zehnder interferometer based spectrometer~\cite{volkov_infrared_1985, pimenov_prb_2005} which allows measurements of amplitude and phase shift of the polarized light in the frequency range 2$\div$20~cm$^{-1}$ . 
In the transmission measurements, we have used a $3\times3\times0.2$ mm$^3$ plane-parallel plate of DyMn$_2$O$_5$ glued to a 3 mm thick Si crystal with Stycast epoxy. 
Both Si substrate and epoxy are transparent and magnetic field insensitive in our range of experimental parameters.  

\begin{figure}
\includegraphics[width=\columnwidth]{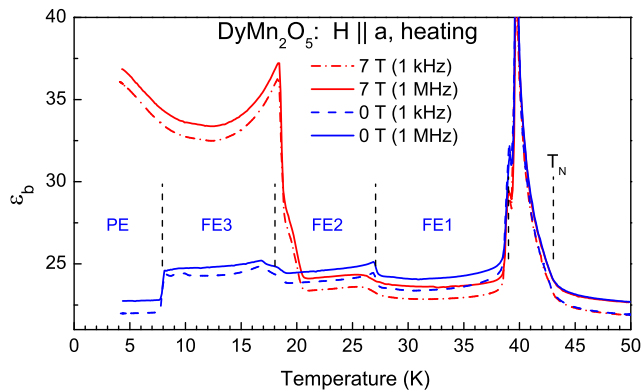}
\caption{(color online). Temperature dependence of the $b$-axis dielectric constant of DyMn$_2$O$_5$ at selected frequencies and magnetic fields. 
Dashed lines separate phases at $H$=0. } 
\label{eps}
\end{figure}

To rule out possible domain-wall-motion origin of the CMD anomaly, measurements of $\varepsilon_b(\omega, T, H_a)$ were performed at frequencies $\omega$ up to 10 MHz, in magnetic field $H_a$ up to 7 T, and at temperatures $T$ below 50~K (Fig.~\ref{eps}). 
The fact that $\varepsilon_b$ in fig.~\ref{eps} does not depend on frequency in the MHz range implies that the frequency of the X-mode is greater than 10 MHz. 
We note that FE relaxor type behavior with characteristic dispersion in kHz range was indeed observed in $R$Mn$_2$O$_5$ multiferroics, albeit above 50 K~\cite{Golovenchits-high-T,Han-JAP-2008}. 

The $H$-$T$ phase diagram of DyMn$_2$O$_5$ was studied by measuring FE and dielectric properties~\cite{Hur-Dy, Higashiyama, Jo}, neutron scattering~\cite{Ratcliff},  magneto-elastic effects~\cite{Cruz2006}, and infrared phonons~\cite{Cao-PRL_2008}. 
At $H$=0, magnetic ordering of Mn ions at 43~K is followed by FE phase transition at 39~K where $\varepsilon_b$ diverges  (Fig.~\ref{eps}). 
At lower temperatures, there are four phases which are characterized by different magnetic orders, electric polarizations and dielectric constants.
Magnetic fields further enrich this phenomenology. 
The CMD effect is observed in a particular phase called FM-IC2 (ferromagnetic incommensurate) above 4 T and below 20~K~\cite{Ratcliff}. 
The shape of $\varepsilon_b$(1~kHz, $T$, 7~T) may be understood as a peak similar to the one at 39~K signaling another FE phase transition, a flat part in the middle similar to Y- and TbMn$_2$O$_5$ step-like anomalies, and an enhancement of the effect below 8~K where Dy ions are fully ordered. 
The CMD effect is maximum at 7~T while at 10 and 17~T the shape of $\varepsilon_b(T)$ is almost a pure step-like anomaly \cite{Jo} very similar to TbMn$_2$O$_5$ at $H$=0. 
 We note that the ionic radius of the Dy ion is between that of Y and Tb suggesting that the unique behavior for the $R$Mn$_2$O$_5$ family must be due to magnetic properties of Dy ions. 
Indeed, ordered magnetic moments of Mn and Dy ions at $H$=0 are oriented mostly along $b$-axis, in contrast with the other $R$Mn$_2$O$_5$ compounds where they are parallel to the $a$-axis.

\begin{figure}
\includegraphics[width=\columnwidth]{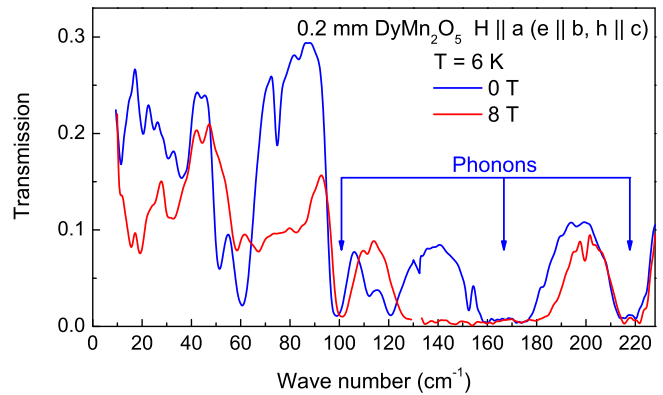}
\caption{(color online). Magnetic field dependence of the far infrared transmission spectrum of the DyMn$_2$O$_5$ single crystal. 
$e$ and $h$ are electric and magnetic fields of light. 
The phonons are identified by insensitivity to magnetic field and by analogy to Y- and TbMn$_2$O$_5$. } 
\label{H}
\end{figure}

We looked for the X-mode in the far infrared frequency range (Fig.~\ref{H}). 
Modes between 30 and 90~cm$^{-1}$ and between 110 and 200~cm$^{-1}$ are electronic $f$-$f$ transitions.
The lowest observed mode in this frequency range is slightly below 20~cm$^{-1}$ and is, most probably, a weak electromagnon doublet such as  was observed in Y- and TbMn$_2$O$_5$~\cite{us-PRL-2007}. 
Interestingly, $f$-$f$ transitions have oscillator strength (assumed to be ED-active) comparable to that of the main TbMn$_2$O$_5$ electromagnon, and they are sensitive to magnetic field.
For DyMn$_2$O$_5$, the total contribution from all modes in this frequency range is $\Delta\varepsilon = \varepsilon(8 T) - \varepsilon(0 T) \approx 1.5$.
Similarly, in measurements of HoMn$_2$O$_5$ Sirenko {\it et al}.~\cite{Sirenko-PRB-2008} found that all modes with frequencies higher than 10~cm$^{-1}$ result in $\Delta\varepsilon = \varepsilon(15~{\text K}, 0~{\text T}) - \varepsilon(25~K, 0~T) \approx 2.5$, which implies the existence of additional mode(s) below 10~cm$^{-1}$. 

\begin{figure}
\includegraphics[width=\columnwidth]{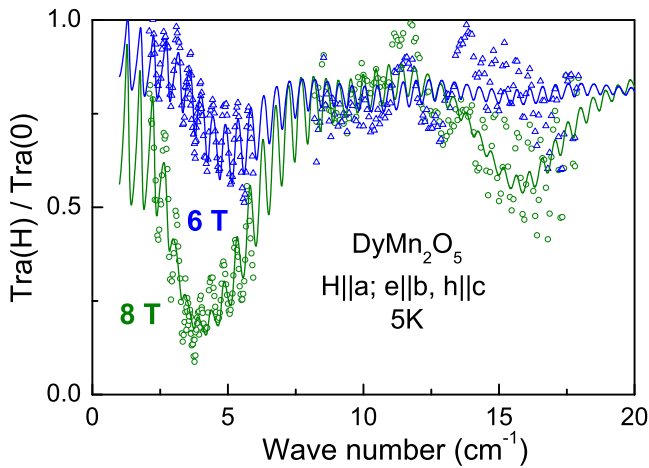}
\caption{(color online). Relative transmission spectra of DyMn$_2$O$_5$ in external magnetic fields.
 Symbols - experiment, solid lines - Lorentzian fit (Eq. (\ref{lorentz}). 
The periodic oscillations in the spectra are due to Fabry-Perot type of resonances (or etalon effect) in the Si substrate which become visible due to the change in the refractive index of DyMn$_2$O$_5$. The mode near 4~cm$^{-1}$ accounts for the CMD effect.}
\label{ftran}
\end{figure}

By extending these measurements to lower frequencies using a set of backward wave oscillators, we then found the putative  low frequency X-mode in DyMn$_2$O$_5$.
Figure~\ref{ftran} shows the ratio of the transmission spectra taken in magnetic field to the zero field spectrum. 
We identify the mode centered near 4~cm$^{-1}$ as the X-mode responsible for the CMD effect. 
This mode is seen only for magnetic fields exceeding 5 T, which agrees with the field-temperature  dependence of the dielectric constant.
The second mode near 16~cm$^{-1}$ is also seen in Fig.~\ref{H}.
We fit the millimeter wave spectra  of DyMn$_2$O$_5$ using the Lorentzian model (Eq.~\ref{lorentz}).
The complex dielectric function $\varepsilon_1+i\varepsilon_2$ of a crystal takes the form:
\begin{equation}\label{lorentz}
    \varepsilon_1+i\varepsilon_2=\varepsilon_{\infty}+
    \sum\frac{\Delta\varepsilon\omega_0^2}{\omega^2-\omega_0^2-i\omega\gamma}
    \quad .
\end{equation}
Here $\Delta\varepsilon$, $\omega_0$, $\gamma$, and $S=\Delta\varepsilon\omega_0^2$ are the dielectric constant contribution, the resonance frequency,  the damping frequency, and the oscillator strength of an ED-active mode, respectively.

\begin{figure}
\includegraphics[width=\columnwidth]{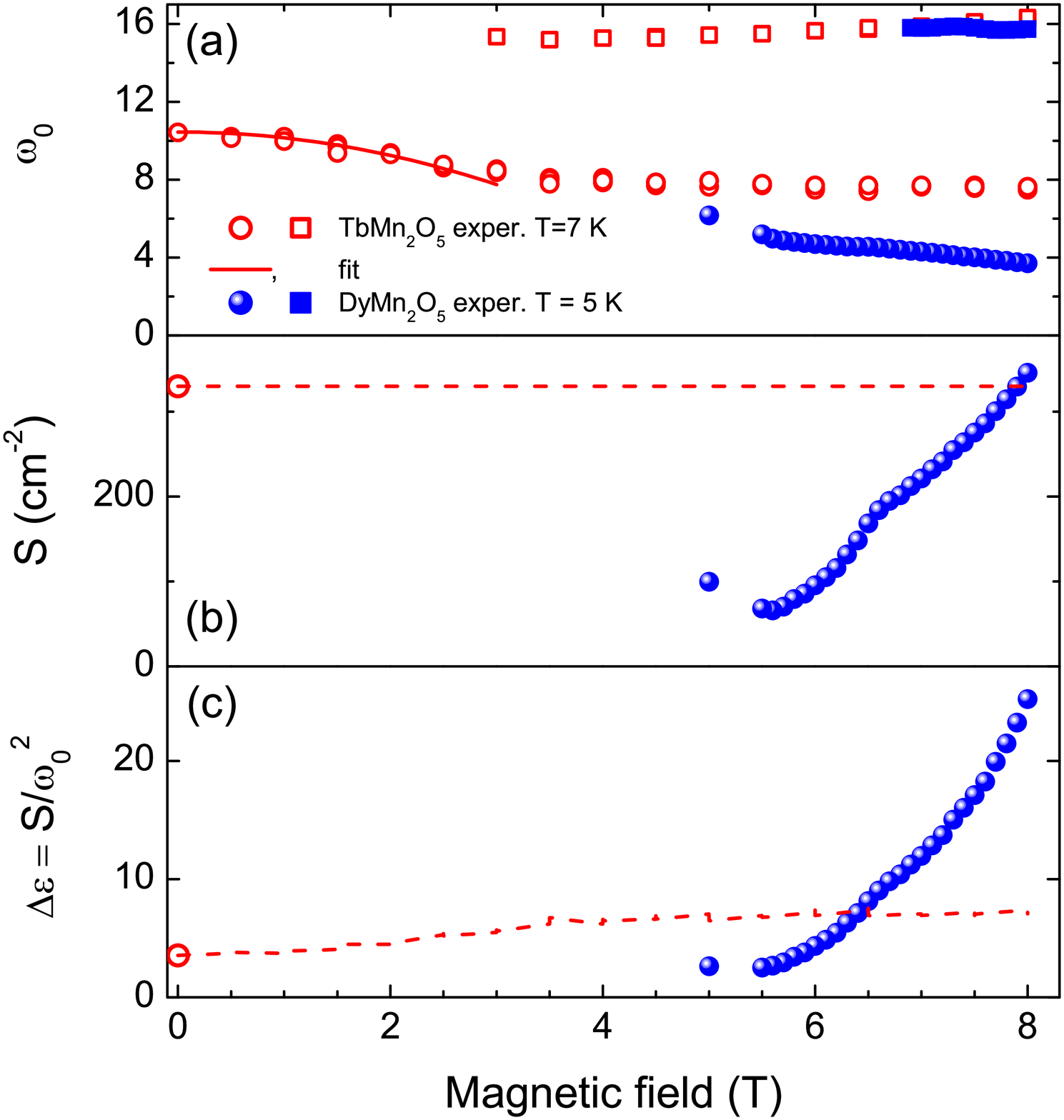}
\caption{(color online). Fitting parameters of the low frequency modes for Tb- and DyMn$_2$O$_5$: (a) resonance frequency, (b) oscillator strength, and (c) contribution to the dielectric constant.
Solid line in (a) is the prediction of Ref.~\onlinecite{Fang-EPL-2008}.
Dashed line in (b) is the electromagnon spectral weight for TbMn$_2$O$_5$ assummed to be field-independent and in (c) it is the calculated contribution of the the low frequency electromagnon to the dielectric constant of TbMn$_2$O$_5$.} 
\label{fitpar}
\end{figure}

The results are shown in Fig.~\ref{fitpar} together with the fitting parameters for TbMn$_2$O$_5$ spectra collected on the  Fourier-transform spectrometer. 
For  TbMn$_2$O$_5$, the resonance frequency of the lowest electromagnon follows the fit curve $\omega_0(H_a) = \sqrt{10.4^2-0.3 H_a^2}$ cm$^{-1}$ in agreement with the prediction of Ref.~\onlinecite{Fang-EPL-2008} up to 3~T.
Above this field, the TbMn$_2$O$_5$ crystal undergoes a transition to another magnetic phase, in which the parameters of this electromagnon are independent of  magnetic field.
This phase extents up to 18~T, and at higher fields the static dielectric constant decreases ~\cite{Haam-FE-2006}, which can be understood as a suppression of the incommensurate magnetic order and electromagnon.
Although we were not able to measure full transmission profile for TbMn$_2$O$_5$, we can conclude that $S$ does not change appreciably with the field and it is assumed to be constant in Fig.~\ref{fitpar}(b). 
Calculated for TbMn$_2$O$_5$ $\Delta \varepsilon(H)$ in panel (c) is in good agreement with the $\varepsilon(H)$ data from the dielectric measurements~\cite{Hur-Dy}. 

The magnetic field behavior of DyMn$_2$O$_5$ is qualitatively different from that of  TbMn$_2$O$_5$ and most of the other $R$Mn$_2$O$_5$ compounds based on available $\varepsilon$(kHz, $T$, $H$) published data. 
The X-mode is observed only at fields higher than 5~T.
There is another phase boundary at 6.5~T, where the field dependence of fit parameters changes.
Both compounds have a higher frequency mode at 16~cm$^{-1}$, albeit in different field ranges: for Tb it is observed above 3 T, while for Dy it appears above 6.5~T. 
The CMD effect is clearly seen in Fig.~\ref{fitpar}(c) and it is caused both by softening of the X-mode  (panel (a)) and by the growth of its spectral weight shown in panel (b). 
Note that at 8~T the X-mode (Fig.~\ref{fitpar}(b)) reaches the spectral weight of the TbMn$_2$O$_5$ electromagnon, which may be evidence in favor of the electromagnon origin of the X-mode.

We now extend our model of ferroelectricity and electromagnons in the 1-2-5 manganites~\cite{us-JPCM-2008, Kim-PRL-2011} to describe the unique  properties of DyMn$_2$O$_5$ (a full theoretical description will be published elsewhere).  
The $R$Mn$_2$O$_5$ consists of two inequivalent antiferromagnetic chains along the $a$ axis formed by Mn spins in the $ab$ layers~\cite{Chapon-YMn2O5}.
Due to the zig-zag shape of the chains the exchange interactions between the chains are frustrated.
They favor an incommensurate spiral state with the approximately right angle between spins in neighboring chains through the Lifshitz invariant, $\bm L_1 \cdot \frac{\partial \bm L_2}{\partial x} - \bm L_2 \cdot \frac{\partial \bm L_1}{\partial x}$, where $\bm L_{1}$($\bm L_{2}$) is the antiferromagnetic order parameter for the chain 1(2).
This state competes with the nearly collinear commensurate state favored by the single-ion magnetic anisotropies. 
The exchange mechanism for magnetoelectric coupling induces an electric polarization along the $b$ axis in the collinear state and makes the oscillations of the angle between the neighboring chains in the spiral state (optical phason) ED active, which enlarges the dielectric constant $\varepsilon_b$ of the spiral state. 
These two effects are mutually exclusive: no electromagnons 
are observed in the collinear FE state, whereas the spiral state with orthogonal spins in neighboring chains shows electromagnon peaks and small electric polarization.

This scenario, common for $R$Mn$_2$O$_5$ compounds, does not work for DyMn$_2$O$_5$. The absence of electromagnons in the incommensurate state of this compound at zero magnetic field can be explained assuming that this state is induced not by the weak Heisenberg interchain interactions, but by antisymmetric Dzyaloshinskii-Moriya (DM) interactions involving the $a$- and $b$-components of spins from neighboring chains, which phenomenologically are described by the Lifshitz invariant, $\left(L_1^x  \frac{\partial L_2^y}{\partial x} - L_2^y \frac{\partial L_1^x}{\partial x}\right) + (1 \leftrightarrow 2)$.
They favor a spiral state in the $ab$-plane and the DM energy is minimized for parallel spins in neighboring chains. 
Thus, if the relativistic interchain interactions are stronger than Heisenberg interactions, an incommensurate state  (FE2 and the states at lower temperatures in Fig.~1) can occur that shows no electromagnons. 
The applied magnetic field along the $a$-axis induces a flop transition, at which the spiral plane changes from $ab$ to $bc$. 
In the $bc$-spiral, the relativistic  interchain interactions become ineffective and the incommensurate state results solely from Heisenberg interactions. 
The angle between spins in neighboring chains then becomes close to 90 degrees, which gives rise to electromagnons and enhanced dielectric susceptibility.
The competition between the spirals with the $0^{\circ}$ and $90^{\circ}$ angles between spins in neighboring chains lowers the frequency of the optical phason, resulting in the CMD effect. 
We note that similar electromagnon must be also present in DyMnO$_3$.
However, the magnetoelectric coupling mechanism for this electromagnon is of relativistic origin~\cite{Katsura-PRL-2007} and its spectral weight is much smaller than that of the exchange electromagnon in DyMn$_2$O$_5$. 
That is why it does not play a role in the DyMnO$_3$ CMD effect.

Another candidate for the X-mode is an ED-active transition between crystal and ligand field split $f$-levels of Dy$^{3+}$ ions.
An optical study of DyAlO$_3$,  where Dy$^{3+}$ ion occupies a site with the same $C_s$ symmetry as in DyMn$_2$O$_5$, revealed a rich manifold of $4f^9$ electron configuration, with an additional (magnetic) 3~cm$^{-1}$ splitting of the ground state below $T_N=3.4$~K, and two metamagnetic transitions in magnetic fields below 2~T~\cite{SCHUCHERT-Z-1969}. 
In magnetic states with broken inversion symmetry, $f$-levels mix with electronic states of opposite parity, which makes the $f$-$f$ transion ED allowed. 
It is not clear, however, whether the ED matrix element between the lowest $f$-levels separated by 4~cm$^{-1}$ in energy is large enough to explain the CMD effect. 

The following arguments are in favor of the electromagnon interpretation: 1) the temperature dependence: the X-mode activates sharply below 20~K, as  is seen from $\varepsilon_b$(kHz,$T$,$H$), while the magnetic ordering of Dy ions is expected only below 8~K; 2) the strength of the X-mode anti-correlates with the electric polarization, as expected for electromagnon \cite{us-JPCM-2008};
3) the CMD effect is observed for electric field strictly along the b-axis, as is the case for electromagnon peaks, while the transitional dipoles for $f$-$f$ transitions are not oriented along crystal axes~\cite{Sirenko-PRB-2008}.
As for the $f$-$f$ transitions, the spectral weight transfer from the ED-allowed electronic transitions is likely to be small due to very small displacements of magnetic ions in 5~T fields and large energy separation between the X-mode and the 2~eV optical gap for the ED-allowed electronic transitions. 
Finally, the crystal and ligand fields in DyMn$_2$O$_5$ are not much different from DyMnO$_3$ where Dy$^{3+}$ ion occupies site of the same local $C_s$ symmetry, but DyMnO$_3$ does not have the X-mode. 

In conclusion, our experimental data and their analysis suggest that the CMD effect in DyMn$_2$O$_5$ is caused by a low-frequency electromagnon (optical phason), which activates in a magnetic-field-induced phase with nearly orthogonal magnetic sublattices.
The strength of this electromagnon is equal to the strength of the electromagnon in TbMn$_2$O$_5$, but its frequency is a factor of two lower, which leads to the four times larger $\varepsilon_b$(kHz, $T$, 7~T) step. 
Further studies, such as theoretical modeling of complex magnetic states and excitations in rare earth manganites with competing interactions between spins,  calculations of the strength of the $f$-$f$ electric dipole transitions, and millimeter wave measurements at higher magnetic fields can lead to better understanding of the magneto-dielectric effects in multiferroics.

\begin{acknowledgments}
This work was supported at Maryland -- by CNAM and NSF MRSEC Grant DMR-0520471, at Vienna -- by the Austrian Science Funds (I815-N16, W1243), at Houston -- by the State of Texas through TCSUH, at Groningen -- by the FOM Grant 08PR2586, at Rutgers -- by the DOE Grant DE-FG02-07ER46382.
\end{acknowledgments} 
%

\end{document}